\documentclass[conference,twoside]{IEEEtran}
\ifCLASSINFOpdf
\else
\fi

\usepackage{amssymb}
\usepackage{amsthm}
\usepackage[cmex10]{amsmath}
\DeclareMathOperator{\E}{\mathbb{E}}

\DeclareMathOperator{\C}{\mathbb{C}}
\DeclareMathOperator{\R}{\mathbb{R}}
\DeclareMathOperator{\Z}{\mathbb{Z}}
\newcommand{\triangleqbin}{\mathbin{\overset\triangle=}}
\newcommand{\mya}{\mathrel{\overset{\makebox[0pt]{{\tiny(a)}}}{=}}}
\newcommand{\myb}{\mathrel{\overset{\makebox[0pt]{{\tiny(b)}}}{=}}}
\newcommand{\myc}{\mathrel{\overset{\makebox[0pt]{{\tiny(c)}}}{=}}}

\usepackage{bm}
\usepackage{fixltx2e}
\usepackage{mathabx}
\usepackage{epsfig,makeidx,color}
\usepackage{graphicx}
\usepackage{amsbsy}
\usepackage{amssymb}
\usepackage{euscript}
\usepackage{lipsum}
\usepackage{cite}
\usepackage{chngcntr}
\usepackage{lineno}
\usepackage[T1]{fontenc}
\usepackage{microtype}
\usepackage{wasysym}
\usepackage[english]{babel}
    \usepackage[font=small,labelsep=space]{caption}
\IEEEoverridecommandlockouts
\begin{document}
%
\title{Energy-Spectral Efficiency Trade-off for a Massive SU-MIMO System with Transceiver Power Consumption}
%
%
%
\author{\IEEEauthorblockN{Sudarshan Mukherjee and Saif Khan Mohammed}
\IEEEauthorblockA{ \thanks{The authors are with the Department of Electrical Engineering, Indian Institute of Technology Delhi (IITD), New delhi, India. Saif Khan Mohammed is also associated with Bharti School of Telecommunication Technology and Management (BSTTM), IIT Delhi. Email: saifkm@iitd.ee.ac.in. This work is supported by EMR funding from the Science and Engineering
Research Board (SERB), Department of Science and Technology (DST),
Government of India.}}
}
\maketitle

\begin{abstract}

We consider a single user (SU) massive MIMO system with multiple antennas at the transmitter (base station) and a single antenna at the user terminal (UT). Taking transceiver power consumption into consideration, for a given spectral efficiency (SE) we maximize the energy efficiency (EE) as a function of the number of base station (BS) antennas $M$, resulting in a closed-form expression for the optimal SE-EE trade-off. It is observed that in contrast to the classical SE-EE trade-off (which considers only the radiated power), with transceiver power consumption taken into account, the EE increases with increasing SE when SE is sufficiently small. Further, for a fixed SE we analyze the impact of varying cell size (i.e., equivalently average channel gain $G_c$) on the optimal EE. We show the interesting result that for sufficiently small $G_c$, the optimal EE decreases as $\mathcal{O}(\sqrt{G_c})$ with decreasing $G_c$. Our analysis also reveals that for sufficiently small SE (or large $G_c$), the EE is insensitive to the power amplifier efficiency.
\end{abstract}


%
\IEEEpeerreviewmaketitle

\section{Introduction}
%
%
%
%
There has been a recent surge of interest on energy efficient ``green communication'' systems, arising out of environmental/cost concerns due to the ever increasing power consumption of cellular systems \cite{Zander}. The total system capacity and energy efficiency is expected to improve significantly in going from the current 4G systems to the next generation cellular communication systems (5G) \cite{Andrews}. Massive MIMO systems has recently been proposed as a possible 5G technology \cite{Marzetta2}. Massive MIMO refers to a communication system where a base station (BS) with $M$ antennas (several tens to hundred) communicates coherently with $K$ user terminals (few tens) on the same time-frequency resource \cite{Marzetta1}.
\par In \cite{Ngo1,Hong} the spectral efficiency (SE) versus energy efficiency (EE) trade-off of massive MIMO system has been studied. However, in these works only the power consumed by the power amplifiers (PA) has been considered. With large $M$ the total power consumed by the $M$ RF transceivers at the BS will becomes significant and must therefore be taken into consideration \cite{Emil4}. The impact of transceiver power consumption on the EE has been studied in several recent papers. However, none of them have derived any closed-form expression for the SE-EE trade-off curve with transceiver power consumption, not even for the single user case with perfect channel state information (CSI). Also none of these recent works have analytically studied the variation in the EE with cell size (for a fixed SE). In the following we briefly discuss the contribution of these recent works.

\par In \cite{Miao} it is shown that the EE of uplink MIMO systems can be optimized by selectively turning off antennas at the user terminal (UT). In\cite{Ha}, the authors optimize the EE of downlink massive MIMO systems with respect to (w.r.t.) the number of BS antennas. It is shown that the EE is a quasi-concave function of the number of BS antennas. In \cite{Hong3} downlink massive MIMO systems are considered, and for a fixed $M$ the EE is maximized w.r.t. the total power radiated from the BS and the number of UTs ($K$). However, results in \cite{Miao, Ha, Hong3} are based on numerical simulations and therefore they provide little insight on the exact SE-EE trade-off. In \cite{Debbah,Emil2} the authors consider the downlink of multi-user MIMO systems, and for the ZF precoder they analytically optimize the EE separately w.r.t. $M$, $K$ and the total radiated power. They show the very interesting result that massive MIMO must be used to increase EE only when interference suppressing multi-user precoding schemes (e.g., ZF, MMSE) are used. However, the analytical results in \cite{Debbah} and \cite{Emil2} cannot be used to derive the exact optimal SE-EE trade-off, since they do not explicitly find the optimal EE for any given SE. In \cite{Mohammed4}, for the uplink multi-user MIMO system with a ZF multi-user detector at the BS, the author studies the impact of varying power consumption parameters (PCPs) on the optimal EE (i.e., EE optimized w.r.t. $(M, K)$ for a given/fixed SE). It is shown that for sufficiently large values of the PCPs it is optimal to have few BS antennas communicating with a single UT, and vice-versa. However since the power consumption model in \cite{Mohammed4} does not consider SE-dependent power consumption due to channel coding/decoding and backhaul, the results in \cite{Mohammed4} cannot be used to derive the optimal SE-EE trade-off.

\par In this paper, we consider the downlink of a single user (SU) system with $M$ antennas at the BS. The UT has a single antenna and the BS is assumed to have perfect CSI.
For this set-up, none of the previous works have derived an analytical expression for the optimal SE-EE trade-off. Similarly no analytical study on the variation in the optimal EE with the cell size (equivalently average channel gain) exists. The main contribution of our paper is the derivation of a closed-form expression for the optimal SE-EE trade-off (for a fixed average channel gain).
We observe that for a sufficiently small SE, the EE increases linearly with SE. This result is in contrast with the classical result where only PA power consumption is taken into consideration, for which the EE always decreases monotonically with increasing SE \cite{Hong}.

\par For a fixed SE, we also analyze the exact variation in the optimal EE with changing average channel gain ($G_c$). It is observed that for sufficiently small $G_c$, with decreasing $G_c$ the EE decreases proportionally to $\sqrt{G_c}$. Through analysis we derive a closed-form expression for the fraction of the total system power consumed by the PAs, as a function of SE (for a fixed $G_c$), and also as a function of $G_c$ (for a fixed SE). It is observed that for sufficiently large SE (or small $G_c$) this fraction is close to half, whereas for sufficiently small SE (or large $G_c$) this fraction is close to zero. Therefore, for sufficiently small SE (or large $G_c$) the EE is insensitive to the PA efficiency. Hence low efficiency PAs (which are generally highly linear \cite{Cripps}) can be used. To the best of our knowledge this study on the variation in the fraction of total power consumed by the PAs as a function of SE (or $G_c$) has not been reported so far in previous works.

\indent \textbf{{Notations:}} $\C$ and $\R$ denote the set of Complex and Real numbers respectively. $\E$ denotes the expectation operator. $(.)^{H}$ denotes the complex conjugate transpose operation. $\Z_{+}$ denotes the set of all positive integers. $\mathcal{C}\mathcal{N}$ denotes the circular symmetric complex Gaussian distribution. Also, $||\bm h||_2 \triangleqbin \left(\sum\limits_{i = 1}^{N}|h_i|^2\right)^{1/2}$ denotes the Euclidean norm of $\bm h = (h_1, h_2, \cdots, h_N) \in \C^{N}$.

\section{System Model}

Let us consider a single user downlink MISO system, where a multiple antenna base station (BS) communicates with a single antenna user terminal (UT). The received signal at the UT is thus given by

\begin{IEEEeqnarray}{c}
\label{eq:system}
y = \sqrt{G_c}\sqrt{P_{\text{T}}}\bm h \bm x + w,
\IEEEeqnarraynumspace
\end{IEEEeqnarray}

\noindent where  $\bm h = (h_1, h_2, \cdots, h_M) \in \C^{1 \times M}$ is the complex baseband channel gain vector, with $h_m$ being the channel gain from the $m^{\text{th}}$ BS antenna to the UT. We assume a Rayleigh flat fading channel, i.e., $h_m \sim \mathcal{C}\mathcal{N}(0,1)$, $m = 1, 2, \ldots, M$.

\noindent Here, $P_{\text{T}}$ is the total transmitted power and $\bm x = (x_1, x_2, \cdots, x_M) \in \C^{M \times 1}$ is the transmitted signal vector ($x_m$ is transmitted from the $m^{\text{th}}$ BS antenna). Let $s \in \C$ be the information symbol to be communicated to the UT ($\E[|s|^2] = 1$). With conjugate beamforming \cite{Hong}, we have\footnote[1]{Perfect CSI at the BS is assumed.}

\begin{IEEEeqnarray}{c}
\label{eq:precode}
\bm x \triangleqbin \dfrac{\bm h^H s}{||\bm h||_2},
\IEEEeqnarraynumspace
\end{IEEEeqnarray}

\noindent In \eqref{eq:system}, $\sqrt{G_c}>0$ models the geometric attenuation and lognormal shadow fading and $w$ represents the additive complex circular symmetric Gaussian white noise at the UT (i.e., $w \sim \mathcal{C}\mathcal{N}(0,N_0B)$). Here $N_0$ is the noise power spectral density and $B$ is the channel bandwidth. The above model can also be applied to wide-band channels, where OFDM is used.

\par The power consumption sources in this model can be categorized as follows: (a) RF power consumption ($P_{\text{RF}}$ in Watt) in the RF chains, power amplifiers (PAs) and oscillator circuits, (b) power consumption ($P_{\text{LP}}$ in Watt) due to conjugate beamforming, (c) fixed power consumption ($P_{\text{s}}$ in Watt) in the baseband processors, and (d) load/data-rate dependent power consumption ($P_{\text{dec}}$ in Watt/bits/s), e.g. channel coding, decoding and backhaul processing. Total load-dependent power consumption is thus computed to be $RBP_{\text{dec}}$, where $R$ is the spectral efficiency (SE) of the system, measured in bits/s/Hz. So, the total system power consumption is given by

\begin{IEEEeqnarray}{c}
\label{eq:totalpower}
P \triangleqbin P_{\text{RF}} + P_{\text{LP}} + P_{\text{s}} + RBP_{\text{dec}}.
\IEEEeqnarraynumspace
\end{IEEEeqnarray}

\indent The RF power consumption can be further subdivided into: (i) per antenna RF chain power consumption at the BS ($P_{\text{BS}}$) and UT ($P_{\text{UT}}$), (ii) power consumption by the local oscillators ($P_{\text{OSC}}$), and (iii) PA power consumption $\alpha P_{\text{T}}$ ($\alpha>1$ models the power efficiency\footnote[2]{Power efficiency of PA is the fraction of input power (consumed power), that is radiated by the antenna \cite{Hong3}. Further we also assume that the PAs operate in the linear region of their transfer characteristics \cite{Mohammed4}.}). Therefore,
\begin{IEEEeqnarray}{c}
\label{eq:rfchain}
P_{\text{RF}} = MP_{\text{BS}} + P_{\text{UT}} + P_{\text{OSC}} + \alpha P_{\text{T}}.
\IEEEeqnarraynumspace
\end{IEEEeqnarray}

\noindent From \eqref{eq:precode}, it is clear that the conjugate beamformer requires a total of $2M$ operations per channel use for scaling and multiplication \cite{Vandenberghe}. Assuming $C_0$ Joule is consumed for each operation, the total energy consumed for $2M$ operations is $2MC_0$ Joule. Since $2M$ operations are performed in ${1}/{B}$ seconds, the overall power consumption for beamforming is $P_{\text{LP}} = 2MC_0B$ Watt. Substituting the expressions for $P_{\text{RF}}$ (from \eqref{eq:rfchain}) and $P_{\text{LP}}$ in \eqref{eq:totalpower}, we get

\begin{IEEEeqnarray}{rCl}
\label{eq:combopower}
\nonumber P & = & M(P_{\text{BS}} + 2C_0B) + P_{\text{UT}} + P_{\text{OSC}} + P_{\text{s}} + RBP_{\text{dec}} + \alpha P_{\text{T}},\\
& = & M(P_{\text{BS}} + 2C_0B) + P_{\text{C}} + RBP_{\text{dec}} + \alpha P_{\text{T}}
\IEEEeqnarraynumspace
\end{IEEEeqnarray}

\noindent where $P_{\text{C}} \triangleqbin P_{\text{UT}} + P_{\text{OSC}} +P_{\text{s}}$.

\par The energy efficiency (EE) in bits/Joule is defined as

\begin{IEEEeqnarray}{c}
\label{eq:eta}
\eta \triangleqbin \dfrac{RB}{P}.
\IEEEeqnarraynumspace
\end{IEEEeqnarray}

\indent Multiplying both sides of \eqref{eq:eta} by $\dfrac{N_0}{G_c}$, we get $\dfrac{\zeta}{R} = \dfrac{N_0B}{G_cP}$, where $\zeta$ is the normalized EE, defined as $\zeta \triangleqbin \eta \dfrac{N_0}{G_c}$. Using \eqref{eq:combopower} we have

%
%
%

\begin{IEEEeqnarray}{c}
\label{eq:eescale}
\dfrac{R}{\zeta} = \dfrac{MG_c}{N_0B} (P_{\text{BS}} + 2C_0B)+ R\dfrac{G_cP_{\text{dec}}}{N_0}+ \dfrac{G_c P_{\text{C}}}{N_0B} + \alpha \dfrac{G_cP_{\text{T}}}{N_0B}.
\IEEEeqnarraynumspace
\end{IEEEeqnarray}

\par Clearly, $\zeta$ is a function of $M$, $R$ and the system parameter vector\footnote[3]{This normalization of the PCPs ($P_{\text{BS}}, C_0B, P_{\text{C}}, BP_{\text{dec}}$) with $N_0B$ is motivated from the fact that power consumption in bandlimited transceiver circuits is typically proportional to $N_0B$ \cite{Mezghani2}.} $\Theta \triangleqbin (\alpha, \rho, \rho_c, \rho_d)$, where

\begin{IEEEeqnarray}{lCl}
\label{eq:sysparam}
\rho \triangleqbin \dfrac{G_c}{N_0B}(P_{\text{BS}} + 2C_0B),\text{      }\rho_c \triangleqbin \dfrac{G_cP_{\text{C}}}{N_0B},\text{      } \rho_d \triangleqbin \dfrac{G_cP_{\text{dec}}}{N_0}.
\IEEEeqnarraynumspace
\end{IEEEeqnarray}

\noindent Substituting \eqref{eq:sysparam} in \eqref{eq:eescale}, $\zeta$ is obtained in terms of $M$, $R$ and $\Theta$ as follows:
\begin{IEEEeqnarray}{c}
\label{eq:invee}
\dfrac{1}{\zeta(M,R,\Theta)} \triangleqbin \rho_d + \dfrac{1}{R}(M\rho + \rho_c) + \alpha \dfrac{\gamma}{R},
\IEEEeqnarraynumspace
\end{IEEEeqnarray}

\noindent where $\gamma \triangleqbin \dfrac{G_c P_{\text{T}}}{N_0B}$ is the transmitted SNR\footnote[4]{$\gamma$ is a function of $M$ and $R$, but we do not write it explicitly for sake of brevity.}.

\section{Energy and Spectral Efficiency Trade-off}

Since we are interested in understanding the EE-SE trade-off, in this section, we study the optimization of $\zeta(M, R, \Theta)$ with respect to the number of BS antennas, $M$, for a fixed $R$ and $\Theta$. The ergodic capacity of the system in \eqref{eq:system} is given by

\begin{IEEEeqnarray}{c}
\label{eq:ergo}
C \triangleqbin \E_h\left[\log_2\left(1 + \gamma ||\bm h||_2^2\right)\right].
\IEEEeqnarraynumspace
\end{IEEEeqnarray}

\indent Clearly, for a given $M$, $C$ increases monotonically with increasing $\gamma$. Therefore, a unique function $\gamma_0(M, R)$ exists such that for any $C' > 0$, $\gamma' = \gamma_0(M, C')$ satisfies the equality $C' = \E_h[\log_2(1 + \gamma'||\bm h||_2^2)]$. With $R = C$, the inverse energy efficiency (from \eqref{eq:invee}) is therefore given by

\begin{IEEEeqnarray}{c}
\label{eq:capee}
\dfrac{1}{\zeta_0(M,R,\Theta)} \triangleqbin \rho_d + \dfrac{1}{R}(M\rho + \rho_c) + \alpha \dfrac{\gamma_0}{R}.
\IEEEeqnarraynumspace
\end{IEEEeqnarray}

\indent For a given $(R, \Theta)$, the exact optimal EE of the system (when the system is capacity achieving, i.e., $R = C$) is given by

\begin{IEEEeqnarray}{c}
\label{eq:optee0}
\nonumber \zeta_0^{\asterisk }(R,\Theta) \triangleqbin  \zeta_0(M_0^{\asterisk}(R,\Theta),R,\Theta),\text{  where}\\
M_0^{\asterisk}(R,\Theta) \triangleqbin \arg \min\limits_{\substack{M \in \Z_{+},\\ M \geq 1}} \dfrac{1}{\zeta_0(M,R,\Theta)}.
\IEEEeqnarraynumspace
\end{IEEEeqnarray}

\indent \textbf{{Lemma 1:}} It can be shown from \eqref{eq:ergo} that

\begin{IEEEeqnarray}{c}
\label{eq:selb}
\log_2(1 + M\gamma) \geq C \geq \log_2(1 + (M-1)\gamma).
\IEEEeqnarraynumspace
\end{IEEEeqnarray}

\begin{IEEEproof}
See Appendix A. \hfill \IEEEQEDhere
\end{IEEEproof}

\indent \textit{{Remark 1:}} It is clear that $\log_2(1 + (M - 1)\gamma)$ is an achievable information rate, lower than the ergodic capacity. With $R = \log_2(1 + (M-1)\gamma)$, for a given $M$ and $R$, the required transmit SNR is given by

\begin{IEEEeqnarray}{c}
\label{eq:snrub}
\gamma_1(M,R) \triangleqbin \dfrac{2^R - 1}{M-1}.
\IEEEeqnarraynumspace
\end{IEEEeqnarray}

\par The corresponding system energy efficiency with $R = \log_2(1 + (M-1)\gamma)$ as the information rate would be


\begin{IEEEeqnarray}{c}
\label{eq:eebound}
\dfrac{1}{\zeta_1(M,R,\Theta)} \triangleqbin \rho_d + \dfrac{1}{R}(M\rho + \rho_c) + \dfrac{\alpha}{R}\dfrac{2^R - 1}{M - 1}.
\IEEEeqnarraynumspace
\end{IEEEeqnarray}
\noindent where $\zeta_1(M, R, \Theta) \triangleqbin \eta_1(M, R, \Theta) N_0/G_c$.\hfill \qed

\par The optimization in \eqref{eq:optee0} is difficult to solve analytically, due to non-availability of closed form expression for $\gamma_0(M, R)$. However, we know that since for a given $M$, $\gamma_0(M, R) \leq \gamma_1(M, R)$, it follows that $\zeta_0(M, R,\Theta) \geq \zeta_1(M, R, \Theta)$. In the following we discuss the tightness of this lower bound on $\zeta_0(M, R, \Theta)$. It follows from Lemma 1 that

\begin{IEEEeqnarray}{c}
\label{eq:remark1}
\gamma_1(M, R) = \dfrac{2^R - 1}{M - 1} \geq \gamma_0(M, R) \geq \dfrac{2^R - 1}{M}.
\IEEEeqnarraynumspace
\end{IEEEeqnarray}

\indent Since, $\dfrac{2^R - 1}{M - 1} \approx \dfrac{2^R - 1}{M}$ for $M \gg 1$ (i.e., for massive MIMO systems), from \eqref{eq:remark1} it follows that $\frac{2^R - 1}{M - 1} \approx \gamma_0(M, R) \approx \frac{2^R - 1}{M}$ for $M \gg 1$. Using this fact and comparing the R.H.S. in \eqref{eq:capee} and \eqref{eq:eebound} we see that for $M \gg 1$, $\zeta_1(M, R, \Theta) \approx \zeta_0(M, R, \Theta)$. The tightness of this approximation has been exhaustively validated through numerical simulations (e.g. see Fig.~\ref{fig:varGc}). We therefore propose to analyze the optimal EE with $R = \log_2(1 + (M - 1)\gamma)$. This optimization is given by
\begin{IEEEeqnarray}{c}
\label{eq:optmee1}
\nonumber M_1^{\asterisk}(R, \Theta) \triangleqbin \arg \min\limits_{\substack{M \in \Z_{+}\\ M \geq 1}} \dfrac{1}{\zeta_1(M, R, \Theta)}, \text{ and}\\
\zeta_1^{\asterisk}(R, \Theta) \triangleqbin \zeta_1(M_1^{\asterisk}(R, \Theta), R, \Theta)
\IEEEeqnarraynumspace
\end{IEEEeqnarray}

\par The optimization problem in \eqref{eq:optmee1} is still difficult to solve in closed form, because $M \in \Z_{+}$. However, a near optimal closed-form solution to the optimization problem in \eqref{eq:optmee1} can be obtained through analysis, if the constraint on $M$ is relaxed so that $M \in \R$. For $M \in \R$ and $M \geq 1$, 

\begin{IEEEeqnarray}{c}
\label{eq:convex}
\dfrac{\partial^ 2}{\partial M^2}\left\{\dfrac{1}{\zeta_1(M,R,\Theta)}\right\}>0, \text{                 } \forall M > 1
\IEEEeqnarraynumspace
\end{IEEEeqnarray}

\noindent Therefore for $M > 1$, $\dfrac{1}{\zeta_1(M, R, \Theta)}$ is convex, with a unique global minimum occurring at

\begin{IEEEeqnarray}{rCl}
\label{eq: Moptima}
M'(R,\Theta) & \triangleqbin & \arg \min\limits_{\substack{M \in \R\\ M \geq 1}} \dfrac{1}{\zeta_1(M, R, \Theta)} = 1 + \sqrt{\dfrac{\alpha}{\rho}(2^R - 1)}.
\IEEEeqnarraynumspace
\end{IEEEeqnarray}

\noindent With $M = M'(R, \Theta)$, the proposed near-optimal energy efficiency is then given by

\begin{IEEEeqnarray}{rCl}
\label{eq:optee}
\nonumber \zeta'(R, \Theta) & \triangleqbin & \zeta_1(M'(R, \Theta), R, \Theta),\\
& = & \dfrac{R}{\rho + \rho_c + R\rho_d + 2\sqrt{\alpha \rho (2^R - 1)}}.
\IEEEeqnarraynumspace
\end{IEEEeqnarray}

\indent Since $\dfrac{1}{\zeta_1(M, R, \Theta)}$ is convex for $M > 1$, it follows that the approximate optimal $M'(R, \Theta)$ and the exact optimal $M_1^{\asterisk}(R, \Theta)$ are close, to be precise, $|M_1^{\asterisk}(R, \Theta) - M'(R, \Theta)| < 1$. Since $M_1^{\ast}(R, \Theta)$ and $M'(R, \Theta)$ are close and $\dfrac{1}{\zeta_1(M, R, \Theta)}$ is continuous in $M$, with bounded derivatives, it is expected that $|\zeta_1^{\asterisk}(R, \Theta) - \zeta'(R, \Theta)| = |\zeta_1(M_1^{\asterisk}(R, \Theta), R, \Theta) - \zeta_1(M'(R, \Theta), R, \Theta)|$ is small, i.e., the relaxation of \eqref{eq:optmee1} to \eqref{eq: Moptima} is near optimal (see also Fig.~\ref{fig:varGc}).




\section{Study of the EE-SE Trade-off}

In this section we study the EE-SE trade-off in \eqref{eq:optee} and characterize its behaviour for small and large values of SE. Throughout this study of the trade-off, we assume $G_c$ and $\Theta$ to be fixed.

\indent \textbf{{Proposition 1:}} For a sufficiently small $R$, i.e., 

\begin{IEEEeqnarray}{c}
\label{eq:inequality1}
R\rho_d + 2\sqrt{\alpha \rho (2^R - 1)} \ll \rho,
\IEEEeqnarraynumspace
\end{IEEEeqnarray}

\noindent we have

\begin{IEEEeqnarray}{rCl}
\label{eq:proposition1}
\zeta'(R, \Theta) & \approx & \dfrac{R}{\rho + \rho_c},
\IEEEeqnarraynumspace
\end{IEEEeqnarray}

\noindent i.e., $\zeta'(R, \Theta)$ increases linearly with increasing $R$. Also for any $R$ satisfying \eqref{eq:inequality1}, we have

\begin{IEEEeqnarray}{c}
\label{eq:Mprop1}
M'(R, \Theta) \approx 1.
\IEEEeqnarraynumspace
\end{IEEEeqnarray}

\begin{IEEEproof}
From \eqref{eq:optee}, it is clear that for $R$ satisfying \eqref{eq:inequality1}, the denominator is dominated by $(\rho + \rho_c)$, from which we get \eqref{eq:proposition1}. Any $R$ satisfying \eqref{eq:inequality1}, also satisfies $2\sqrt{\alpha \rho (2^R - 1)} \ll \rho$, i.e., $\sqrt{\dfrac{\alpha}{\rho}(2^R - 1)} \ll \dfrac{1}{2}$, using which in \eqref{eq: Moptima}, we get \eqref{eq:Mprop1}. \hfill \IEEEQEDhere
\end{IEEEproof}

\indent \textit{{Remark 2:}} In the following we explain the result of proposition 1. When $R$ is sufficiently small, the required power to be radiated from the PAs is also small. Therefore the total system power consumption is dominated by the power consumed by the RF chains at the BS and UT and the fixed power consumption (e.g. oscillators and baseband processors etc.). With increase in $M$, the power consumed by the RF chains at the BS will increase and therefore EE will decrease. This shows that for sufficiently small $R$, it is optimal to have only a single BS antenna (i.e., non-massive MIMO regime, see \eqref{eq:Mprop1}).

\par From the above discussion we know that with a sufficiently small $R$, it is expected that the PA power consumption is a small fraction of the total system power consumption. Since the number of BS antennas is fixed to one, the total system power consumption will be almost constant with increasing $R$ (as long as $R$ is sufficiently small). This implies that the overall EE will increase with increasing $R$ (see Fig.~\ref{fig:varR}). This result is in contrast with the classical result where the EE always decreases with increasing SE when only PA power consumption is considered.  \hfill \qed


\indent \textbf{{Proposition 2:}} For a sufficiently large $R$, i.e., 

\begin{IEEEeqnarray}{c}
\label{eq:inequality2}
R\rho_d + 2\sqrt{\alpha \rho (2^R - 1)} \gg \rho + \rho_c
\IEEEeqnarraynumspace
\end{IEEEeqnarray}

\noindent $\zeta'(R, \Theta)$ decreases monotonically with increasing $R$ and that $\lim\limits_{R \to \infty} \zeta'(R, \Theta) = 0$.

\indent Further $M'(R, \Theta)$ increases exponentially with increasing $R$.


\begin{IEEEproof}
For any $R$ satisfying \eqref{eq:inequality2}, the denominator of the R.H.S. of \eqref{eq:optee} is dominated by $R\rho_d + 2\sqrt{\alpha \rho (2^R - 1)}$. Therefore under the condition in \eqref{eq:inequality2}, we get

\begin{IEEEeqnarray}{c}
\label{eq:largeR}
\zeta'(R, \Theta) \approx \dfrac{1}{\rho_d + 2\sqrt{\alpha \rho \left(\dfrac{2^R - 1}{R^2}\right)}}
\IEEEeqnarraynumspace
\end{IEEEeqnarray}

\indent From \eqref{eq:largeR} it is clear that $\zeta'(R, \Theta)$ decreases monotonically with increasing $R$ (sufficiently large) and that $\lim\limits_{R \to \infty}\zeta'(R, \Theta) = 0$. Further, from \eqref{eq: Moptima} it follows that $M'(R, \Theta)$ increases exponentially with $R$ as $R \to \infty$. \hfill \IEEEQEDhere

\end{IEEEproof}

\indent \textit{{Remark 3:}} As $R \to \infty$, the power consumed by the PAs, increase at a rate proportional to $2^R$ for a fixed $M$ (see \eqref{eq:snrub}). Increasing $M$ will increase the array gain, which in turn will reduce the PA power consumption. If $M \, \propto \, 2^{R/2 + \epsilon}$ ($\epsilon > 0$), then $M$ increases at a rate faster than $2^{R/2}$ and the total power consumption is dominated by the RF chain power consumption at the BS (see \eqref{eq:eebound}), which increases linearly with $M$. Similarly if $M \, \propto \, 2^{R/2 - \epsilon}$, i.e., $M$ increases at a rate slower than $2^{R/2}$, the total system power consumption is dominated by the power consumed by the PA, which increases as $2^{R/2 + \epsilon}$. In either case, the total power consumption increases as $2^{R/2 + \epsilon}$. Therefore, it is optimal to have $\epsilon=0$, i.e., $M \, \propto \, 2^{R/2}$ (massive MIMO regime). \hfill \qed

\begin{figure*}[!hbt]
\normalsize
\begin{IEEEeqnarray}{lCl}
\eta'(R,\Theta) \triangleqbin \zeta'(R, \Theta)\dfrac{G_c}{N_0} = \dfrac{RB}{P_{\text{BS}} + 2C_0B + P_{\text{C}} + RBP_{\text{dec}} + 2\sqrt{\dfrac{N_0B}{G_c}}\sqrt{\alpha(2^R - 1)(P_{\text{BS}} + 2C_0B)}}.
\label{eq:EEnoscale}
\end{IEEEeqnarray}
\hrulefill
\vspace*{4pt}
\end{figure*}


\section{Impact of Cell Size on EE (Fixed SE and PCPs)}

In this section we analyze the impact of $G_c$ (i.e., average channel gain) on the EE of the system for a fixed $R$ and fixed power consumption parameters $(\alpha, P_{\text{dec}}, P_{\text{BS}}, C_0, P_{\text{C}})$. We assume that the UT is at the cell edge and therefore a large value of $G_c$ corresponds to a small cell size and vice versa. Since $\Theta = (\alpha, \rho, \rho_c, \rho_d)$ depends on $G_c$, using \eqref{eq:optee} the unnormalized near optimal EE is given by \eqref{eq:EEnoscale}. Similarly from \eqref{eq: Moptima} the near optimal $M$ is given by

\begin{IEEEeqnarray}{c}
\label{eq:Mnoscale}
M'(R, \Theta) = 1 + \sqrt{\dfrac{N_0B}{G_c}}\sqrt{\alpha\left(\dfrac{2^R - 1}{P_{\text{BS}} + 2C_0B}\right)}.
\IEEEeqnarraynumspace
\end{IEEEeqnarray}

\indent \textbf{{Proposition 3:}} If $G_c$ is sufficiently large, i.e.,

\begin{IEEEeqnarray}{c}
\label{eq:inequality3}
2\sqrt{\dfrac{N_0B}{G_c}}\sqrt{\alpha (2^R - 1)(P_{\text{BS}} + 2C_0B)} \ll (P_{\text{BS}} + 2C_0B).
\IEEEeqnarraynumspace
\end{IEEEeqnarray}

\noindent then the near optimal  EE, $\eta'(R, \Theta)$ becomes insensitive to changes in $G_c$. Further, for any $G_c$ satisfying \eqref{eq:inequality3}, we have $M'(R, \Theta) \approx 1$.

\begin{IEEEproof}
From \eqref{eq:EEnoscale} it is evident that for any $G_c$ satisfying \eqref{eq:inequality3}, the total power consumption would be dominated by $(P_{\text{BS}} + 2C_0B + P_{\text{C}} + RBP_{\text{dec}})$. Using \eqref{eq:inequality3} in \eqref{eq:EEnoscale}, we therefore have

\begin{IEEEeqnarray}{c}
\label{eq:eeprop3}
\eta'(R, \Theta) \approx \dfrac{RB}{P_{\text{BS}} + 2C_0B + P_{\text{C}} +RBP_{\text{dec}}}.
\IEEEeqnarraynumspace
\end{IEEEeqnarray}

\indent Clearly, the R.H.S. of \eqref{eq:eeprop3} is not a function of $G_c$ and therefore $\eta'(R, \Theta)$ is insensitive to changes in $G_c$. From \eqref{eq:inequality3} it follows that $\sqrt{\frac{N_0B}{G_c}}\sqrt{\frac{{\alpha(2^R - 1)}}{(P_{\text{BS}} + 2C_0B)}} \ll \frac{1}{2}$. Using this in \eqref{eq:Mnoscale} we get $M'(R, \Theta) \approx 1$. \hfill  \IEEEQEDhere
\end{IEEEproof}

\indent \textit{{Remark 4:}} For small cell size, the effective channel gain $G_c$ could be large, resulting in reduction in the required transmit power. Thus the power consumed by the PAs will decrease with increasing $G_c$ and the total system power consumption will eventually be dominated by the other sources of power consumption (including the power consumed by the $M$ RF antennas at the BS).

\par With increase in $M$, the power consumed by the RF chains at the BS will increase and dominate the total system power consumption. Hence with increasing $M$ the EE will decrease. This shows that for sufficiently large $G_c$ (i.e. for small cell size) it is optimal to have only a single antenna at the BS.

\par
\begin{figure}[t]
\hspace{-0.5 in}
\includegraphics[width= 4.2 in, height= 2.8 in]{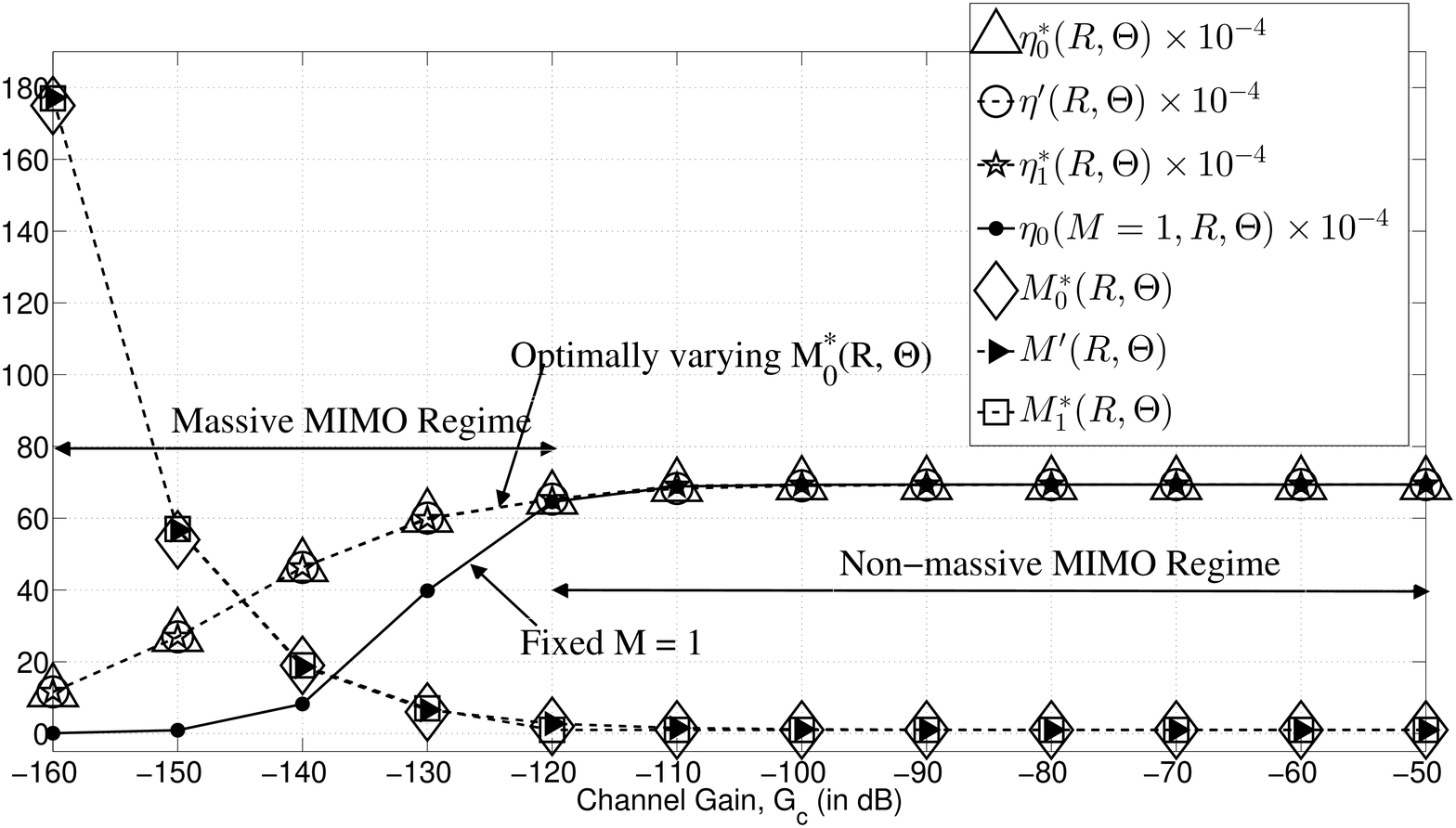}
\caption {Unnormalized Optimal $\eta_0^{\ast}(R, \Theta)$ and $M_0^{\ast}(R, \Theta)$ versus $G_c$ for a fixed spectral efficiency $R = 5$ bps/Hz and fixed system parameters: $N_0 = 10^{-20.4}$ W/Hz, $P_{\text{s}} = 5$ W, $P_{\text{dec}} = 1.15$ W/Gbits/s, $P_{\text{BS}} = P_{\text{UT}} = 0.1$ W, $P_{\text{OSC}} = 2$ W, $B = 1$ MHz, PA power efficiency = $1/\alpha = 0.39$, and $C_0 = 10^{-9}$ J.}
\label{fig:varGc}
\end{figure}

 With a single antenna at the BS, the total system power consumption is almost constant with increasing $G_c$. This is so because with increasing $G_c$ the power consumption by PAs is increasingly dominated by the other sources of power consumption (which are independent of $G_c$). Hence for a fixed SE, the EE is insensitive to changes in the cell size as long as the cell size is sufficiently small (see also the non-massive MIMO regime in Fig.~\ref{fig:varGc}). \hfill \qed

\indent \textbf{{Proposition 4:}} If $G_c$ is sufficiently small, i.e.,

\begin{IEEEeqnarray}{rCl}
\label{eq:inequality4}
\nonumber 2\sqrt{\dfrac{N_0B}{G_c}}\sqrt{\alpha (2^R - 1)(P_{\text{BS}} + 2C_0B)} & \gg & P_{\text{BS}} + 2C_0B\\
 & + & RBP_{\text{dec}} + P_{\text{C}},
\IEEEeqnarraynumspace
\end{IEEEeqnarray}

\noindent then $\eta'(R, \Theta) \, \propto \, \sqrt{G_c}$ with decreasing $G_c$. Furthermore, $M'(R, \Theta) \, \propto \, 1/\sqrt{G_c}$, i.e., $M'(R, \Theta)$ increases with decreasing $G_c$.

\begin{IEEEproof}
It is clear that for small $G_c$ satisfying \eqref{eq:inequality4}, the denominator of \eqref{eq:EEnoscale} is dominated by the L.H.S. of \eqref{eq:inequality4}. Using this fact in \eqref{eq:EEnoscale}, we get

\begin{IEEEeqnarray}{c}
\label{eq:eeprop4}
\eta'(R, \Theta) \approx \sqrt{G_c} \left(\dfrac{R}{2\sqrt{\dfrac{N_0}{B}}\sqrt{\alpha (2^R - 1)(P_{\text{BS}} + 2C_0B)}}\right),
\IEEEeqnarraynumspace
\end{IEEEeqnarray}

\indent i.e., $\eta'(R, \Theta) \, \propto\, \sqrt{G_c}$. Further from \eqref{eq:Mnoscale}, it is evident that $M'(R, \Theta) \, \propto \, {1}/{\sqrt{G_c}}$ as $G_c \to 0$. \hfill \IEEEQEDhere
\end{IEEEproof}

\indent \textit{{Remark 5:}} For fixed $M$ and $R$, the required radiated power from the BS increases linearly with decreasing $G_c$. Therefore for sufficiently small $G_c$, the PA power consumption will dominate the total power consumption. Hence with a fixed $M$, the energy efficiency would decrease linearly with decreasing $G_c$. By increasing $M$, we can reduce the PA power consumption due to increase in array gain. However, increasing $M$ will also increase the RF power consumption. It follows that the best trade-off is observed by increasing $M \, \propto \, {1}/{\sqrt{G_c}}$ with decreasing $G_c$ (i.e., massive MIMO regime). With $M \, \propto \, {1}/{\sqrt{G_c}}$, both the PA and the RF power consumptions will increase at the same rate (i.e. $\, \propto \, {1}/{\sqrt{G_c}}$) with decreasing $G_c$. Therefore, as $G_c$ (sufficiently small) decreases, the overall EE will decrease as\footnote[5]{This same conclusion has also been drawn in \cite{Mohammed4} through heuristic arguments but no analytical proof has been presented.} $\sqrt{G_c}$. Note that by varying $M \, \propto \, {1}/{\sqrt{G_c}}$ with decreasing $G_c$, the EE reduces at a slower rate (i.e., $\, \propto \, \sqrt{G_c}$) compared to a linear decrease for the scenario where $M$ is fixed (see also Fig.~\ref{fig:varGc}). \hfill \qed

\section{Effect of SE and $G_c$ on the PA Design}

In this section we analyze the impact of SE and $G_c$ on the required power efficiency of the PAs at the BS. We observe that for sufficiently large SE or large cell size the PAs must be highly efficient. In contrast to this, for sufficiently small SE or small cell size the EE is insensitive to the PA efficiency and therefore low efficiency PAs can be used. We firstly compute the fraction of the total system power consumption that is consumed by the PAs. The total system power consumption with near optimal $M$ is given by the denominator of the R.H.S. in \eqref{eq:EEnoscale}, i.e.

\begin{IEEEeqnarray}{rCl}
\label{eq:totpower}
\nonumber P & = & P_{\text{BS}} + 2C_0B + P_{\text{C}} + RBP_{\text{dec}}\\
&& \> + 2\sqrt{\dfrac{N_0B}{G_c}}\sqrt{\alpha(2^R - 1)(P_{\text{BS}} + 2C_0B)}.
\IEEEeqnarraynumspace
\end{IEEEeqnarray}

\noindent Similarly, with near optimal $M$, the power consumption by the PAs is given by

\begin{IEEEeqnarray}{rCl}
\label{eq:papower}
\nonumber \alpha P_{\text{T}} & \mya  & \dfrac{\alpha N_0B}{G_c}\gamma_1(M'(R, \Theta), R) \myb \dfrac{\alpha N_0B}{G_c}\dfrac{2^R - 1}{M'(R, \Theta) - 1}\\
\nonumber & \myc & \sqrt{\dfrac{N_0B}{G_c}}\sqrt{\alpha (2^R - 1) (P_{\text{BS}} + 2C_0B)}
\IEEEeqnarraynumspace
\end{IEEEeqnarray}

\noindent where $(a)$ follows from \eqref{eq:invee}, $(b)$ follows from \eqref{eq:snrub}, and $(c)$ follows from \eqref{eq:Mnoscale}. The fraction of the total power consumed by the PAs, $f_{\text{PA}} \triangleqbin \dfrac{\alpha P_{\text{T}}}{P}$ is therefore given by

\begin{IEEEeqnarray}{rCl}
\label{eq:fraction}
f_{\text{PA}} & =  \left[2 + \sqrt{G_c}\dfrac{(P_{\text{BS}} + 2C_0B + P_{\text{C}} + RBP_{\text{dec}})}{\sqrt{N_0B}\sqrt{\alpha (2^R - 1)}}\right]^{-1}
\IEEEeqnarraynumspace
\end{IEEEeqnarray}

\begin{figure}[t]
\hspace{-0.2 in} \includegraphics[width= 4.0 in, height= 2.5 in]{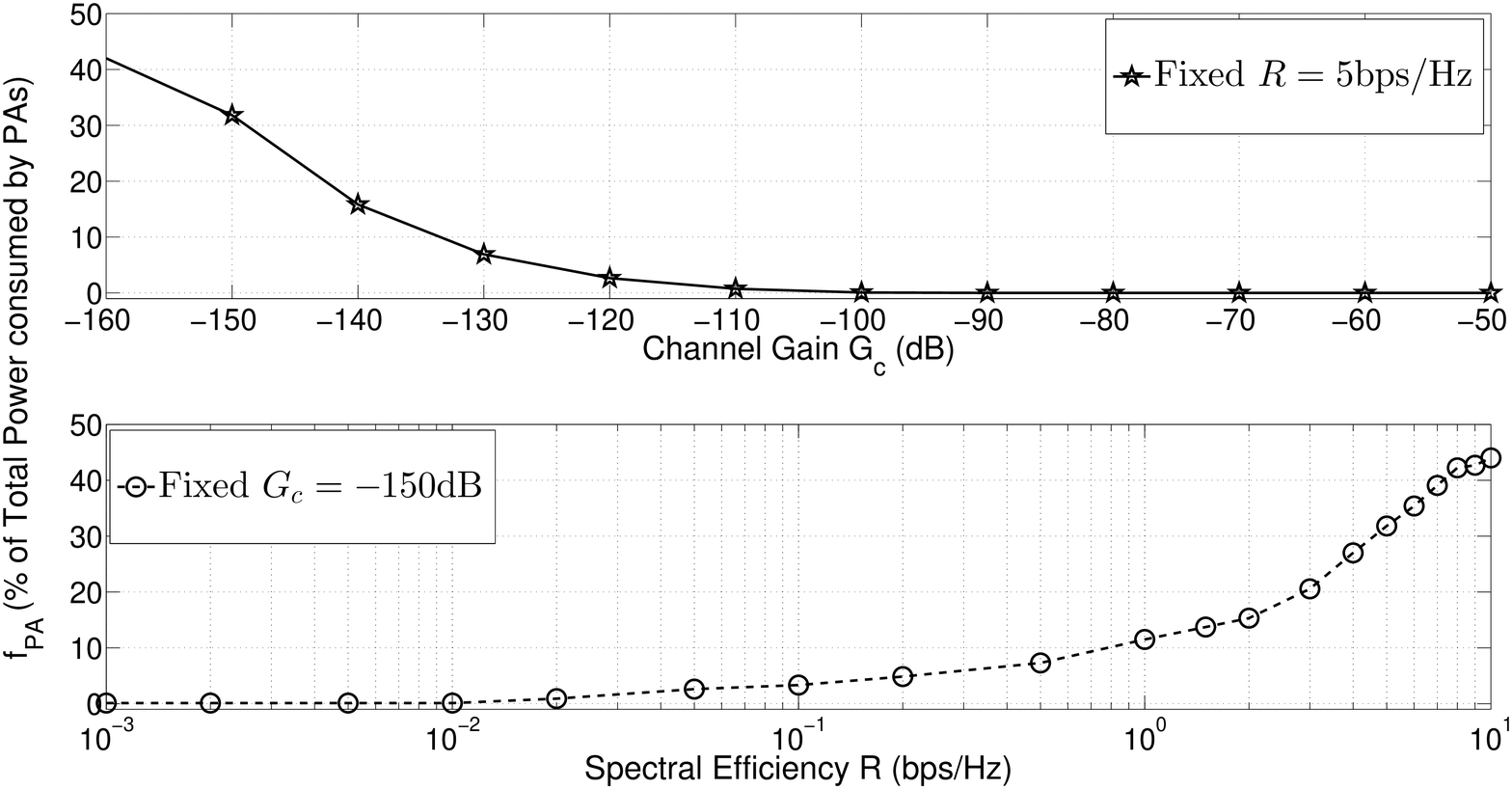}
\caption {Fraction of the total system power consumed by the PAs as a function of (i) Varying $R$ and fixed channel gain $G_c$, and (ii) varying channel gain $G_c$ and fixed $R$.}
\label{fig:varGcRalpha}
\end{figure}

\indent \textit{{Remark 6:}} For sufficiently small $G_c$ (i.e. large cell size) or for sufficiently large $R$, from \eqref{eq:fraction} it is clear that $f_{\text{PA}} \approx \frac{1}{2}$ (with $G_c \to 0$ or $R \to \infty$, see also Fig.~\ref{fig:varGcRalpha}). This result is expected since for large cell size or large $R$, the total power consumption is equally dominated by the RF power consumption at the BS (including beamformer) and the PA power consumption. Therefore in the massive MIMO regime highly efficient PAs must be used. \hfill \qed

\indent \textit{{Remark 7:}} From \eqref{eq:fraction} it is clear that with $R \to 0$ or $G_c \to \infty$ (i.e. small SE or small cell size) we have $f_{\text{PA}} \to 0$. From Remark 2 and Remark 4 we know that for small SE or small cell size it is EE optimal to operate in the non-massive MIMO regime, i.e., $M'(R, \Theta) \approx 1$. With $M'(R, \Theta) \approx 1$, the power consumption by the RF chain is constant. However as $R \to 0$ or $G_c \to \infty$, the PA power consumption goes to zero. Hence the fraction of total power consumed by the PAs is negligible as $R \to 0$ or $G_c \to \infty$. This leads us to the conclusion that the EE is insensitive to the PA power efficiency as $R \to 0$ or $G_c \to \infty$ (see also Fig.~\ref{fig:varGcRalpha}). Hence in the non-massive MIMO regime low efficiency PAs can be used.\hfill \qed

\begin{figure}[t]
\hspace{-0.3 in} 
\includegraphics[width= 4.0 in, height= 2.8 in]{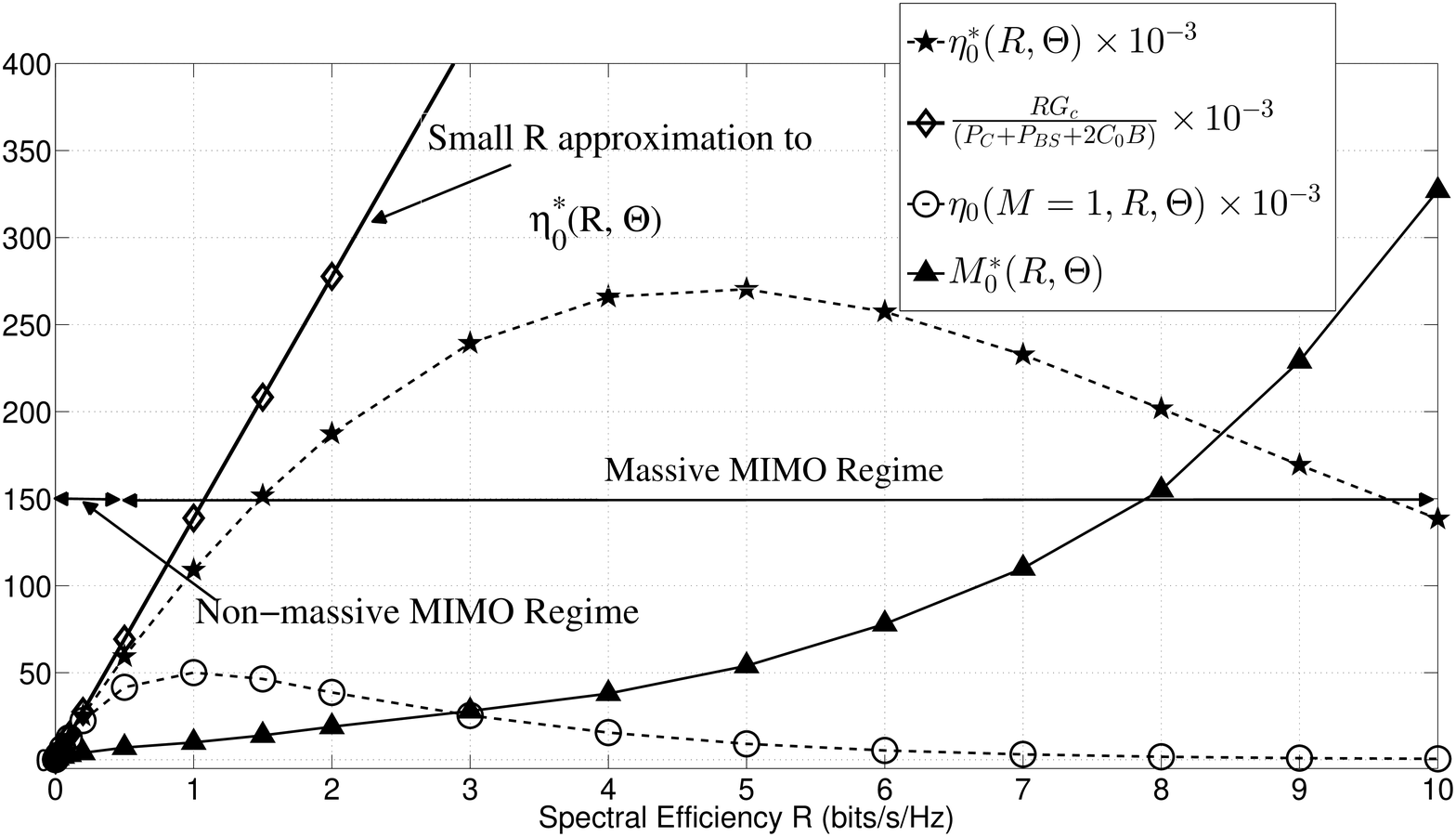}
\caption {Unnormalized Optimal $\eta_0^{\ast}(R, \Theta)$ and $M_0^{\ast}(R, \Theta)$ versus $R$ for a fixed channel gain $G_c = -150$ dB and fixed system parameters: $N_0 = 10^{-20.4}$ W/Hz, $P_{\text{s}} = 5$ W, $P_{\text{dec}} = 1.15$ W/Gbits/s, $P_{\text{BS}} = P_{\text{UT}} = 0.1$ W, $P_{\text{OSC}} = 2$ W, $B = 1$ MHz, PA power efficiency = $1/\alpha = 0.39$, and $C_0 = 10^{-9}$ J.}
\label{fig:varR}
\end{figure}

\section{Numerical Results}

In this section, we numerically compute the exact optimal EE $\eta_0^{\ast}(R, \Theta) \triangleqbin \zeta_0^{\ast}(R, \Theta)\dfrac{G_c}{N_0}$ and the exact optimal $M_0^{\ast}(R, \Theta)$. We present simulation curves in support of the analysis done in sections III to VI. Throughout this section, we assume $N_0 = 10^{-20.4}$ W/Hz, $P_{\text{s}} = 5$ W, $P_{\text{dec}} = 1.15$ W/Gbits/s, $P_{\text{BS}} = P_{\text{UT}} = 0.1$ W, $P_{\text{OSC}} = 2$ W, $B = 1$ MHz, PA power efficiency = $1/\alpha = 0.39$, and $C_0 = 10^{-9}$ J. These are based on realistic values obtained from prior works \cite{Auer, Mezghani, Kumar, greenlist}.

\par In Fig.~\ref{fig:varGc} we study the impact of $G_c$ on the EE of the system for a fixed SE ($R = 5$ bps/Hz). We observe that it is EE optimal to have a single antenna at the BS when $G_c$ is sufficiently large ($G_c > -120$ dB). In this non-massive MIMO regime, the EE is observed to be almost constant with varying $G_c$. These observations support the analytical findings of Proposition 3 (see also Remark 4).

\par With decreasing $G_c$ ($G_c$ is sufficiently small), it is observed that the EE decreases and the optimal number of BS antennas increases (i.e. massive MIMO regime). We have also plotted the EE, $\eta_0(M = 1, R, \Theta) \triangleqbin \zeta_0(M = 1, R, \Theta){G_c}/{N_0}$, achieved when the number of BS antenna $M$ is fixed to one (i.e., $M = 1$ does not vary with $G_c$). It is observed that $\eta_0(M = 1, R, \Theta)$ decreases at a much faster rate compared to the optimal EE $\eta_0^{\ast}(R, \Theta)$ (see Remark 5). For instance at $G_c = -140$ dB we have $\frac{\eta_0^{\ast}(R = 5, \Theta)}{\eta_0( M = 1, R = 5, \Theta)} \approx 5.65$ and at $G_c = -150$ dB we have $\frac{\eta_0^{\ast}(R = 5, \Theta)}{\eta_0( M = 1, R = 5, \Theta)} \approx 29.59$. Further, in Fig.~\ref{fig:varGc} it can be seen that the proposed near optimal EE in \eqref{eq:EEnoscale} is close  to the exact optimal EE, i.e., $\eta'(R, \Theta) \approx \eta_0^{\ast}(R, \Theta)$. This justifies the discussion in section III on the near optimal relaxation of the exact EE optimization in \eqref{eq:optee0} by the optimization in \eqref{eq: Moptima}.


\par In Fig.~\ref{fig:varR}, we plot the exact optimal EE-SE trade-off for a fixed $G_c = -150$ dB. It is observed that the EE increases linearly with $R$ for sufficiently small $R$ (compare the curve with black stars to that with diamonds). This observation supports Proposition 1 and Remark 2. With increasing $R$, the EE decreases eventually as is suggested in Proposition 2 and Remark 3.

\par In Fig.~\ref{fig:varGcRalpha} we plot the fraction of the total system power consumed by PAs ($f_{\text{PA}}$) as a function of : (i) varying $G_c$, with fixed $R = 5$ bits/s/Hz and (ii) varying $R$ with fixed $G_c = -150$ dB. For sufficiently small $R$ or large $G_c$, it is observed that $f_{\text{PA}}$ is small and therefore low efficiency PAs can be used. In contrast for the massive MIMO regime (i.e. sufficiently large $R$ or small $G_c$), we observe that almost $50\%$ of the total system power consumption is due to the PAs and therefore power efficient PAs must be used. These observations support the analytical findings in Remarks 6 and 7.

%



%

\appendices
\section{Proof of Lemma 1}
Since $||\bm h||_2^2$ is $\chi^2$ distributed with $2M$ degrees of freedom, we have
\begin{IEEEeqnarray}{c}
\label{eq:expect}
\E_h\left[||\bm h||_2^2\right] = M, \text{  }\E_h\left[\dfrac{1}{||\bm h||_2^2}\right] = \dfrac{1}{M - 1}
\IEEEeqnarraynumspace
\end{IEEEeqnarray}

\indent Since $f(x) \triangleqbin \log_2(1 + x)$ is concave in $x$, from Jensen's inequality it follows that
\begin{IEEEeqnarray}{rCl}
\nonumber \E_h[\log_2(1 + \gamma||\bm h||_2^2)] & \leq & \log_2(1 + \gamma \E_h[||\bm h||_2^2])\\
& = & \log_2(1 + M\gamma).
\IEEEeqnarraynumspace
\end{IEEEeqnarray}
\noindent where the last step follows from \eqref{eq:expect}. Similarly, $g(x) \triangleqbin \log_2(1 + \frac{1}{x})$ is convex in $x$, and therefore from Jensen's inequality we have

\begin{IEEEeqnarray}{rCl}
\nonumber \E_h[\log_2(1 + \gamma||\bm h||_2^2)] & = & \E_h\left[\log_2\left(1 + \dfrac{\gamma}{{1}/{||\bm h||_2^2}}\right)\right]\\
\nonumber & \geq & \log_2\left(1 + \dfrac{\gamma}{{\E_h\left[\dfrac{1}{||\bm h||_2^2}\right]}}\right)\\
& = & \log_2(1 + (M - 1)\gamma).
\IEEEeqnarraynumspace
\end{IEEEeqnarray}

%


\ifCLASSOPTIONcaptionsoff
  \newpage
\fi



%

\bibliographystyle{IEEEtran}
\bibliography{IEEEabrvn,mybibn}

\end{document}